\documentclass[10pt,twocolumn,twoside]{IEEEtran}
\usepackage{amsmath,amsfonts}
\usepackage{algorithmic}
\usepackage{algorithm}
\usepackage{array}
\usepackage{textcomp}
\usepackage{stfloats}
\usepackage{url}
\usepackage{verbatim}
\usepackage{graphicx}
\usepackage{cite}
\usepackage{amssymb}
\usepackage{bm}%
\usepackage{times}
\usepackage{subfigure}
\usepackage{enumitem}
\usepackage{latexsym}
\usepackage{hhline}
\usepackage{caption}
\usepackage{algorithm} 
\usepackage{algorithmic} 

\usepackage{multirow} 
\usepackage{xcolor}
\usepackage{epstopdf}
\usepackage{aligned-overset}

\begin{document}

    \title{Towards Standardizing OTFS: A Candidate Waveform for Next-Generation Wireless Networks}

\author{\IEEEauthorblockN{
Mingcheng Nie, Ruoxi Chong,
Shuangyang Li,~\IEEEmembership{Member,~IEEE,}
Arman Farhang,~\IEEEmembership{Senior Member,~IEEE,}\\
Fabian Göttsch,
Derrick Wing Kwan Ng,~\IEEEmembership{Fellow,~IEEE},\\
Michail Matthaiou,~\IEEEmembership{Fellow,~IEEE},
and
Yonghui Li,~\IEEEmembership{Fellow,~IEEE}\\
\emph{(Invited Paper)}
}

\thanks{The work of M. Nie is supported by the University of Sydney International Stipend Scholarship.
This work was supported by the U.K. Engineering and Physical Sciences Research Council (EPSRC) grant (EP/X04047X/2) for TITAN Telecoms Hub.
The work of R. Chong and M. Matthaiou was supported by the European Research Council (ERC) under the European Union’s Horizon 2020 research and innovation programme (grant agreement No. 101001331).
The work of S. Li was supported in part by European Union’s Horizon 2020 Research and Innovation Program through MSCA under Grant 101105732-DDComRad. The work of A. Farhang was supported by Taighde Éireann - Research Ireland under Grant numbers 21/US/3757 and 23/ARC/12082.
\emph{(Corresponding author: Ruoxi Chong.)}}

\thanks{M. Nie and Y. Li are with the School of Electrical and Computer Engineering, University of Sydney, Sydney NSW 2006, Australia (e-mail: \{mingcheng.nie, yonghui.li\}@sydney.edu.au).}

\thanks{R. Chong and M. Matthaiou are with the Centre for Wireless Innovation (CWI), Queen’s University Belfast, U.K. (e-mail: \{rchong02, m.matthaiou\}@qub.ac.uk).}

\thanks{S. Li is with the Department of Electrical Engineering and Computer Science, Technische Universit{\"a}t Berlin, 10587 Berlin, Germany (e-mail: shuangyang.li@tu-berlin.de).}

\thanks{A. Farhang is with the Department of Electronic, Electrical Engineering, Trinity College Dublin, D02 PN40 Dublin, Ireland (e-mail: arman.farhang@tcd.ie).}

\thanks{F. Göttsch is with Massive Beams GmbH, 10625 Berlin, Germany (email: fabian.goettsch@massivebeams.com).}

\thanks{D. W. K. Ng is with the School of Electrical Engineering and Telecommunications, University of New South Wales, Sydney, NSW 2052, Australia (email: w.k.ng@unsw.edu.au).}
}



\maketitle

\begin{abstract}
The standardization of the sixth-generation (6G) has recently commenced to address the rapidly growing demands for enhanced wireless network services. Nevertheless, existing wireless systems, particularly at the physical layer waveform level, remain inadequate for achieving the ambitious key performance indicators (KPIs) envisioned for 6G.  Specifically, orthogonal frequency division multiplexing (OFDM), the widely adopted waveform in fifth-generation new radio (5G-NR) networks, suffers from inherent limitations in satisfying these stringent requirements. In practice, OFDM can experience severe inter-carrier interference (ICI), resulting in a pronounced data rate error floor caused by high Doppler shifts. Additionally, the repetitive usage of cyclic prefixes (CPs), intended to combat multipath delays, results in significant spectral inefficiency. These fundamental drawbacks pose critical obstacles to fulfilling 6G performance objectives. Orthogonal time frequency space (OTFS) modulation has recently emerged as a promising waveform candidate, addressing the aforementioned challenges by exploiting the unique characteristics of the delay-Doppler (DD) domain channel. Unlike OFDM, OTFS is inherently resilient to channel distortions induced by delay and Doppler effects, while remaining sensitive to time and frequency shifts. Such intrinsic properties are instrumental in enabling OTFS, with joint communication and sensing capabilities, to embrace, rather than combat, dynamic channel conditions. Motivated by these compelling advantages, this article investigates the feasibility and practical implementation of OTFS modulation leveraging the current OFDM-based wireless systems. Furthermore, we present a practical precoding scheme for OTFS-based multiple-input multiple-output (MIMO) systems, characterized by near-linear computational complexity and compared with conventional OFDM-based MIMO systems. We also highlight the advantages of the DD waveform for integrated sensing and communications (ISAC). Through these explorations, we reveal that the DD waveform has substantial potential to achieve scalable, efficient, and robust wireless communication and sensing capabilities, positioning it as a critical technology for future 6G networks.
\end{abstract}

\section{Introduction}
\subsection{Background}
Wireless communication has become an indispensable enabler for driving technological advancements, with a transformative impact clearly evident in how society connects and exchanges information.  From the low-quality analog voice calls of first-generation (1G) systems in the 1980s to today's ultra-fast and seamlessly interconnected fifth-generation new radio (5G-NR) networks, mobile data traffic has experienced exponential growth, as evidenced by a $25\%$ increase in global mobile data traffic between 2023 and 2024 alone~\cite{Ericsson2025}. This rapid upward trend is expected to continue, with global data traffic projected to reach an astonishing 473 million terabytes per month by the end of 2030~\cite{Ericsson2025}. Furthermore, the number of mobile subscribers is anticipated to rise to 6.5 billion, representing approximately $76\%$ of the global population~\cite{GSMA2024}. Motivated by the escalating demands for efficient, reliable, and ubiquitous information exchange across heterogeneous environments, the era of sixth-generation (6G) networks is rapidly approaching.

\begin{figure}
    \centering
    \includegraphics[scale=0.65]{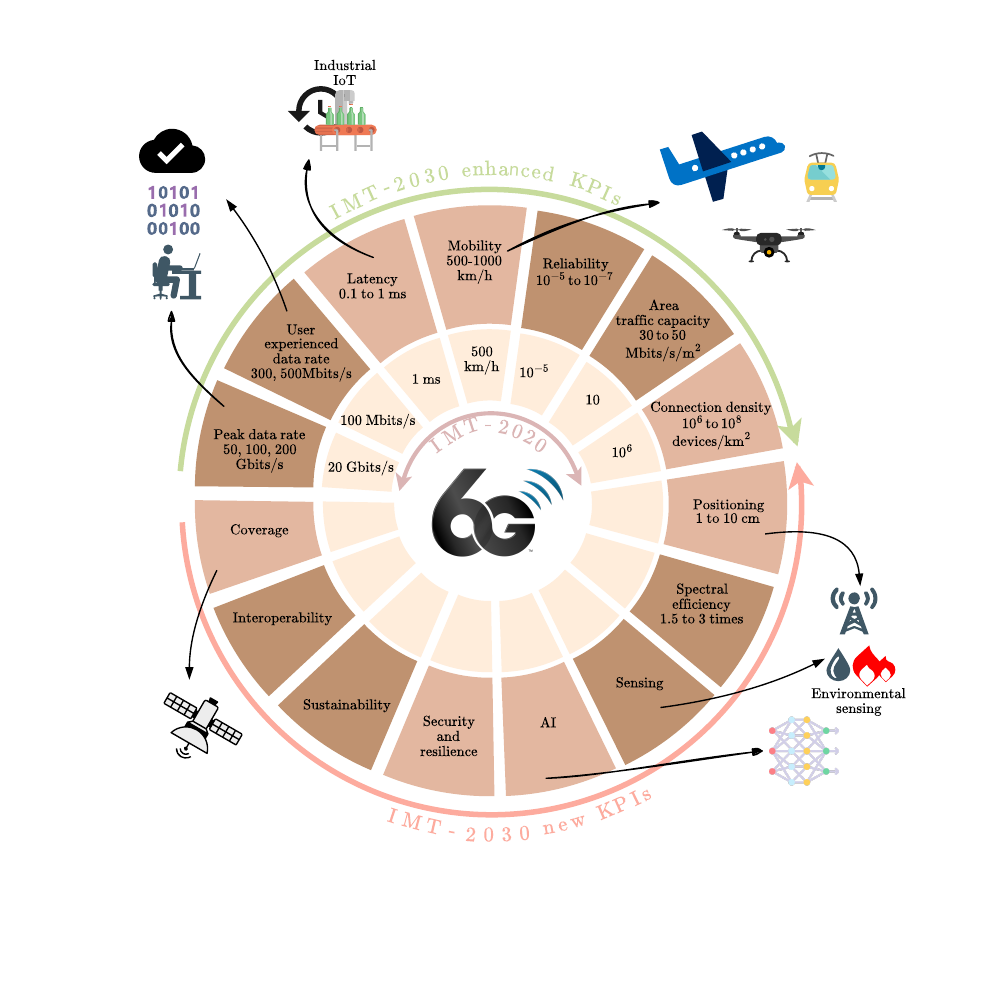}
    \caption{5G-NR (inner circle) versus 6G (outer circle) capabilities~\cite{ITU2023}.}
    \label{fig:6G_KPIs}
\end{figure}

Beyond delivering higher data rates, 6G networks are envisioned to maintain consistently enhanced user experience across highly dynamic conditions, through resilient and ubiquitous connectivity along with integrated sensing capabilities. To realize this vision, the key performance indicators (KPIs) for 6G networks have been outlined in the official International Mobile Telecommunications 2030  (IMT-2030) recommendation~\cite{ITU2023}. As shown in Fig.~\ref{fig:6G_KPIs}, the targeted peak data rate reaches up to $200$ Gbits/s, while maintaining an ubiquitously stable data rate of up to $500$ Mbits/s. In addition, 6G aims to achieve ultra-high reliability, characterized by an error probability below  $10^{-7}$ and latency less than $1$ millisecond. Moreover, spectral efficiency (SE) is targeted to be improved by approximately $1.5$ to $3$ times compared to existing 5G-NR networks, along with the capability of supporting high-accuracy sensing functionalities. Furthermore, a key requirement for achieving true ubiquitous connectivity is to ensure robust and seamless services in challenging environments. These include high-mobility scenarios commonly encountered on highways, high-speed trains, unmanned aerial vehicles (UAVs), aircrafts, autonomous driving platforms, and in the integration of terrestrial and non-terrestrial networks, as illustrated in Fig.~\ref{fig:6G_KPIs}.  To this end, 6G is designed to support user mobility at speeds of up to $1000$ km/h.

However, existing wireless systems are currently insufficient for achieving the stringent KPIs of 6G. To bridge this gap, the development of an advanced physical layer waveform is essential, which needs to  satisfy the following properties: 
\begin{itemize}
    \item Balanced performance between communications and sensing, which enables robust communications over challenging environments, while availing of accurate sensing capabilities;
    
    \item Seamless integration with multiple-input multiple-output (MIMO) configurations, while maintaining acceptable computational complexity to substantially enhance the achievable data rates;
    \item Backward compatibility with current fourth-generation (4G) and 5G-NR networks to minimize deployment challenges, economic costs, and promote the sustainability of 6G networks.
\end{itemize}

Addressing these critical requirements demands significant innovation in waveform design, as current mainstream waveforms are evidently insufficient to fulfill these demands~\cite{wang2006performance,wei2021orthogonal}.

\subsection{Limitations of Time-Frequency Waveform}
Orthogonal frequency division multiplexing (OFDM) is a foundational time-frequency (TF) domain waveform, extensively employed in current 5G-NR networks. By partitioning the available spectrum into multiple overlapped narrowband subcarriers without mutual interference, OFDM enables efficient parallel data transmission, thereby reducing implementation complexity while significantly improving SE. Moreover, the requirement of compatibility between the upcoming 6G technology and existing 5G-NR systems makes the continued use of OFDM technically competitive and economically advantageous. However, despite its witnessed success in 5G-NR developments, OFDM suffers from inherent limitations in coping with the \emph{Doppler effect} caused by the relative motion between transceiver pairs. In particular, Doppler shifts disrupt the orthogonality among sub-carriers, resulting in severe inter-carrier interference (ICI), and consequently, significant error performance degradation. As reported in~\cite{wang2006performance}, conventional OFDM systems suffer from a substantial error floor even at typical highway speeds of $50$ km/h, whereas 6G KPIs demand robust and reliable transmission at user mobility speeds of up to $1000$ km/h. 
To mitigate ICI, the network providers usually increase the subcarrier spacing of OFDM according to the current 5G-NR standard. While this approach alleviates ICI under limited low-mobility scenarios, it comes at the expense of reduced SE, lower achievable data rates, and inter-numerology interference. Furthermore, OFDM suffers from inherent spectral inefficiency due to the repeated use of cyclic prefixes (CPs) to combat multipath delay, resulting in the loss of approximately $7\%$ to $25\%$ of valuable resources wasted for interference mitigation rather than actual data bearing. In addition to communication limitations, the key radar parameters, e.g., delay and Doppler shifts, are not explicitly represented in the conventional TF domain channel model. As a result, additional extraction of sensing parameters from the received signal is required for OFDM sensing, resulting in significant sensing overhead and increased computational cost. Those limitations render the OFDM waveform insufficient to fulfill the stringent performance criteria envisioned for 6G waveforms. Moreover, in MIMO systems, rapid channel variations exacerbate the channel aging effect, which requires frequent channel estimation and precoder updates. This leads to increased pilot overhead and computational burden for OFDM-based MIMO systems, further highlighting OFDM's incompatibility with the required efficiency of MIMO transmission in the 6G era. Consequently, while OFDM remains the dominant waveform in 5G-NR networks, its inherent limitations in handling high-mobility environments, sensing integration, and large-scale MIMO render it inadequate for satisfying the ambitious performance targets set by 6G. Therefore, novel breakthroughs in wireless waveform design are urgently needed.


\subsection{Delay-Doppler Waveform: A Promising Solution}
The representation of the wireless channel is important to the waveform design. Among various modeling approaches, characterizing the channel based on its delay and Doppler response provides a physically intuitive and propagation-aware representation.  While the delay-Doppler (DD) channel model has been extensively utilized in radar sensing applications, it has recently garnered growing attention due to its superior capability to ensure reliable communication links under highly dynamic conditions. In particular, the orthogonal time-frequency space (OTFS) waveform~\cite{hadani2017orthogonal}, as an effective realization of DD waveforms, has emerged as a promising candidate for delivering high data rates in high-mobility scenarios. Moreover, OTFS multiplexes data symbols in the DD domain, which naturally aligns with key sensing parameters, such as delay and Doppler spread, offering significant potential for underpinning integrated sensing and communication (ISAC) functionality. Specifically, OTFS avails of channel delay and Doppler characteristics by ensuring that the signal transmitted over any single resolvable path is shifted in time and frequency but not distorted. In other words, OTFS is inherently resilient to channel distortions induced by delay and Doppler effects, while remaining sensitive to time and frequency shifts caused by physical scatterers. The DD domain's time-invariant representation of the rapidly time-varying TF channel enables the full potential of channel exploitation for both communication and sensing. 
Furthermore, thanks to the quasi-static nature of the DD domain channel, OTFS inherently mitigates the detrimental effects of channel aging, making it highly compatible with MIMO configurations. In particular, the requirement for frequent precoder updates is substantially reduced. 
On top of these benefits, only a single CP is typically required for an OTFS frame. 
This significantly improves SE compared to conventional OFDM waveforms, perfectly aligning with the high-throughput objectives envisioned for 6G wireless networks.

By exploiting the unique properties of the DD domain channel, several practical designs of OTFS have recently been investigated. For example, efficient channel estimation in the DD domain was achieved by comparing the received pilot symbols against a threshold~\cite{hong2022delay}. A sufficiently large guard space was also introduced to prevent interference between pilot and information symbols at the receiver, thereby enabling reliable and efficient transmission~\cite{hong2022delay}. A low-complexity receiver leveraging a classic message passing algorithm was also proposed in~\cite{hong2022delay}, where the DD domain symbol interference was approximated as a Gaussian variable to reduce the computational complexity. The OTFS performance in terms of both sensing and communication was studied in~\cite{gaudio2020effectiveness}, providing a comprehensive comparison with the conventional OFDM waveforms. The results demonstrated that OTFS achieves sensing accuracy comparable to state-of-the-art radar waveforms, such as frequency-modulated
continuous wave (FMCW), while outperforming OFDM in terms of communication throughput.

Despite the aforementioned advantages, the standardization process of OTFS is still in its infancy. For a new waveform to be successfully standardized, ensuring robust backward compatibility with existing systems is essential. Therefore, this article addresses several challenges pertaining to OTFS standardization, highlighting potential solutions for achieving backward compatibility with existing 5G waveforms.

\subsection{Contributions}
In this article, we first provide a comprehensive overview of channel characteristics in the DD domain to unveil the fundamental principles underlying the unique advantages of OTFS. Subsequently, practical implementation strategies for OTFS, based on existing 5G-NR wireless systems, are discussed, demonstrating its feasibility and compatibility for deployment within future 6G standards. Furthermore, we present a reduced-complexity precoder scheme tailored for OTFS-based MIMO, highlighting that OTFS can enable efficient and scalable MIMO transmission. The benefits of performing ISAC in the DD domain are also emphasized by establishing the direct connections between the DD domain channel and the range-Doppler matrix, the latter being widely adopted in sensing applications. Finally, we summarize various key technical challenges and opportunities associated with DD waveform design.

\section{Preliminaries of DD Domain Channels}
\begin{figure}[t]
	\centering  
    \subfigure[An illustration of stationarity and coherence regions: $\mathcal{R}_{c}$ and $\mathcal{R}^{'}_{c}$ denote two consecutive coherence regions, within which the TF channel responses are distinct, i.e., $\mathcal{R}_{c}(t,f)\neq \mathcal{R}^{'}_{c}(t,f)$. A similar principle applies to stationarity region $\mathcal{R}_{s}$ and $\mathcal{R}^{'}_{s}$. ]{{\includegraphics[scale=0.7]{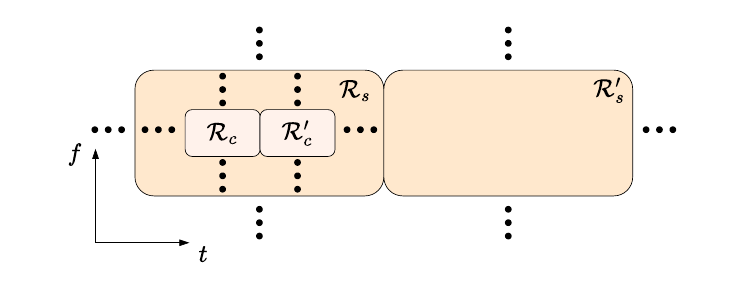}}}
	\subfigure[TF domain channel response]{{\includegraphics[scale=0.29]{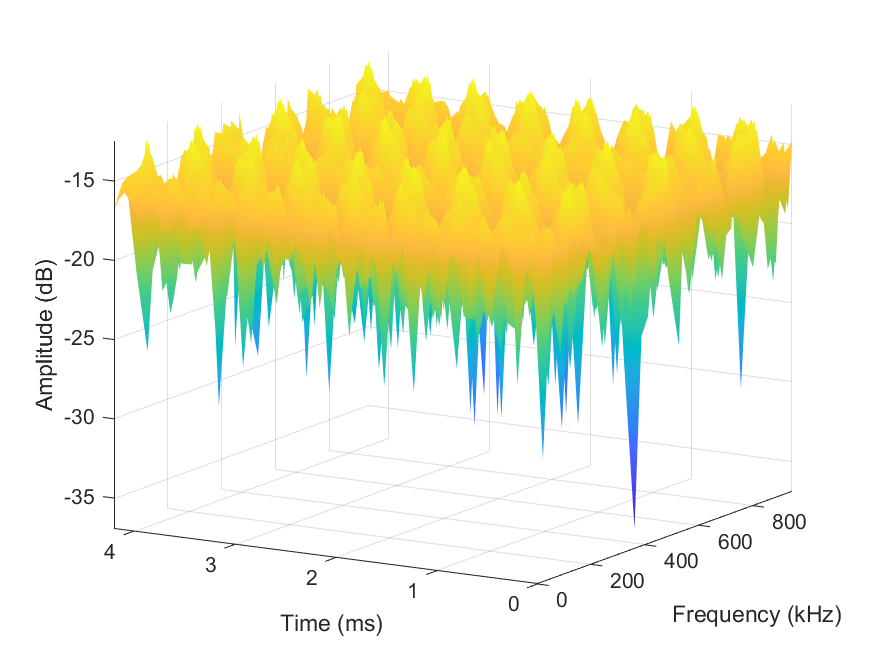}}}
    \subfigure[DD domain channel response]{{\includegraphics[scale=0.29]{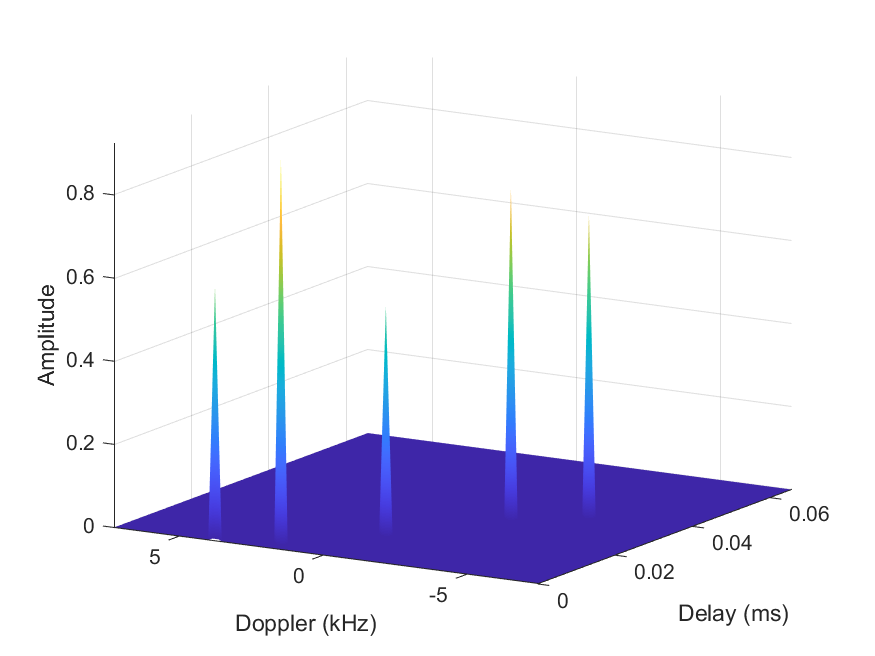}}}
    \vspace{-3mm}
    \caption{WSSUS regions and channel responses in different domains.} 
        \label{fig:TF_DD_channels}
\end{figure}

Conventionally, wireless channels have been studied in the TF domain, where the coherence region is defined as the range of TF components over which the underlying channel response remains relatively stable. An intuitive illustration of the coherence region is depicted in  Fig. \ref{fig:TF_DD_channels}(a), where the channel response remains approximately constant within each coherence region $\mathcal{R}_{c}$, while variations may occur across different coherence regions.
In fact, the size of the coherence region is governed by the underlying physical channel characteristics, i.e., it is inversely proportional to the maximum delay and Doppler spread. As a result, in rapidly varying channels with large Doppler or delay spreads, the coherence region shrinks significantly. This leads to a more pronounced linear time-varying (LTV) behavior, as illustrated in Fig.~\ref{fig:TF_DD_channels}(b), where the channel response varies significantly across different TF locations. Such an LTV channel accurately represents real-world wireless transmission environments and the development of realistic models for these scenarios is essential for the design and evaluation of communication systems. To achieve tractable models, the assumption of wide-sense stationary uncorrelated scatterers (WSSUS) is usually invoked, indicating that the channel's delay and Doppler characteristics are approximately unchanged. While this preserves simplicity for system design, real-world channels typically deviate somewhat from the identical WSSUS assumption~\cite{hlawatsch2011wireless}. Nevertheless, it has been established in~\cite{hlawatsch2011wireless} that even for more general non-WSSUS channels, there exists a stationarity region $\mathcal{R}_{s}$ (cf. Fig.~\ref{fig:TF_DD_channels}(a)), within which the wireless channel can be effectively approximated as WSSUS. As a result, the DD domain channel response remains approximately constant within this region due to the WSSUS assumption, as shown in Fig.~\ref{fig:TF_DD_channels}(c). Physically, this phenomenon can be interpreted as the delay and Doppler associated with the physical scatterers remaining unchanged, provided that the channel geometry remains static during the transmission snapshot. More importantly, the size of this stationarity region is inversely proportional to the maximum delay and Doppler lags, which determines a broad range of TF components over which the channel dynamics are highly correlated. As reported in~\cite{wei2021orthogonal}, DD domain channels can achieve a stationarity time of 100 ms and a stationarity bandwidth of 100 MHz, which is significantly larger than the conventional coherence region. This important property enables the DD domain channel response to remain stable over a wide range of symbols, and facilitates the straightforward extension of the DD waveform to MIMO systems.  {Note that the stationarity region is channel characteristics rather than a quantity dictated by the system parameters, i.e., bandwidth and time duration. Therefore, system parameters can be appropriately chosen so that one or multiple OTFS frames are transmitted within a stationarity region, thereby fully exploiting channel stationarity.}

On top of the aforementioned advantages, the DD domain channel response provides a direct physical interpretation of the propagation environment. Specifically, the DD domain channel is represented by a set of discrete impulses, each associated with a specific delay $\tau$ and Doppler frequency $\nu$, which can be mathematically expressed as
\begin{align}
    h_{\mathrm{DD}}=\sum_{p=0}^{P-1} h_p \delta(\tau-\tau_p)\delta(\nu-\nu_p),
\end{align}
where $P$ denotes the number of distinct physical scatterers and $h_p$, $\tau_p$, and $\nu_p$ denote the fading coefficient, delay, and Doppler shifts of the $p$-th scatterer, respectively. This concise representation directly captures the fundamental channel attributes and offers several attractive advantages:
\begin{itemize}
    \item {\bf Sparsity:} Since the number of dominant physical scatterers in realistic propagation environments is generally limited, the DD domain channel exhibits a naturally sparse structure, characterized by only a few significant components located at specific delays and Doppler frequencies, e.g., Fig.~\ref{fig:TF_DD_channels}(c).
    \item {\bf{Quasi-static:}} The DD domain channel can be viewed as a ``snapshot'' of the propagation environment.  As the characteristics of physical scatterers remain roughly unchanged over the transmission ``snapshot'', the DD domain representation of the channel is effectively quasi-static.
    \item {\bf Separability:} Any paths that are resolvable by the system cannot exhibit identical delay, Doppler, and angular responses simultaneously. Such paths originate from sufficiently separated physical scatterers, rendering them individually distinguishable and resolvable in the DD domain.
    \item {\bf Compactness:} The delay spread and Doppler frequency spread are bounded by the maximum delay and maximum Doppler frequency, respectively, which are physically determined by the maximum propagation distance and relative speed in the wireless environment. As a result, the DD domain channel response is confined within a clearly defined region, yielding a compact and well-structured representation.
\end{itemize}
Furthermore, the DD domain representation offers a significantly finer-grained characterization of channel attributes compared to the conventional TF domain approach~\cite{hlawatsch2011wireless}. Specifically, the resolutions achievable along the delay and Doppler dimensions are inversely proportional to the signal bandwidth and the observation time, respectively, both of which are typically much smaller than the time and frequency resolutions in the TF domain, determined by the symbol duration and subcarrier spacing. As a result, the DD domain representation provides a more precise and physically meaningful description of the channel, enabling waveform and transceiver designs to fully avail of the underlying channel dynamics.

\begin{figure*}[t!]
    \centering
    \includegraphics[scale=0.55]{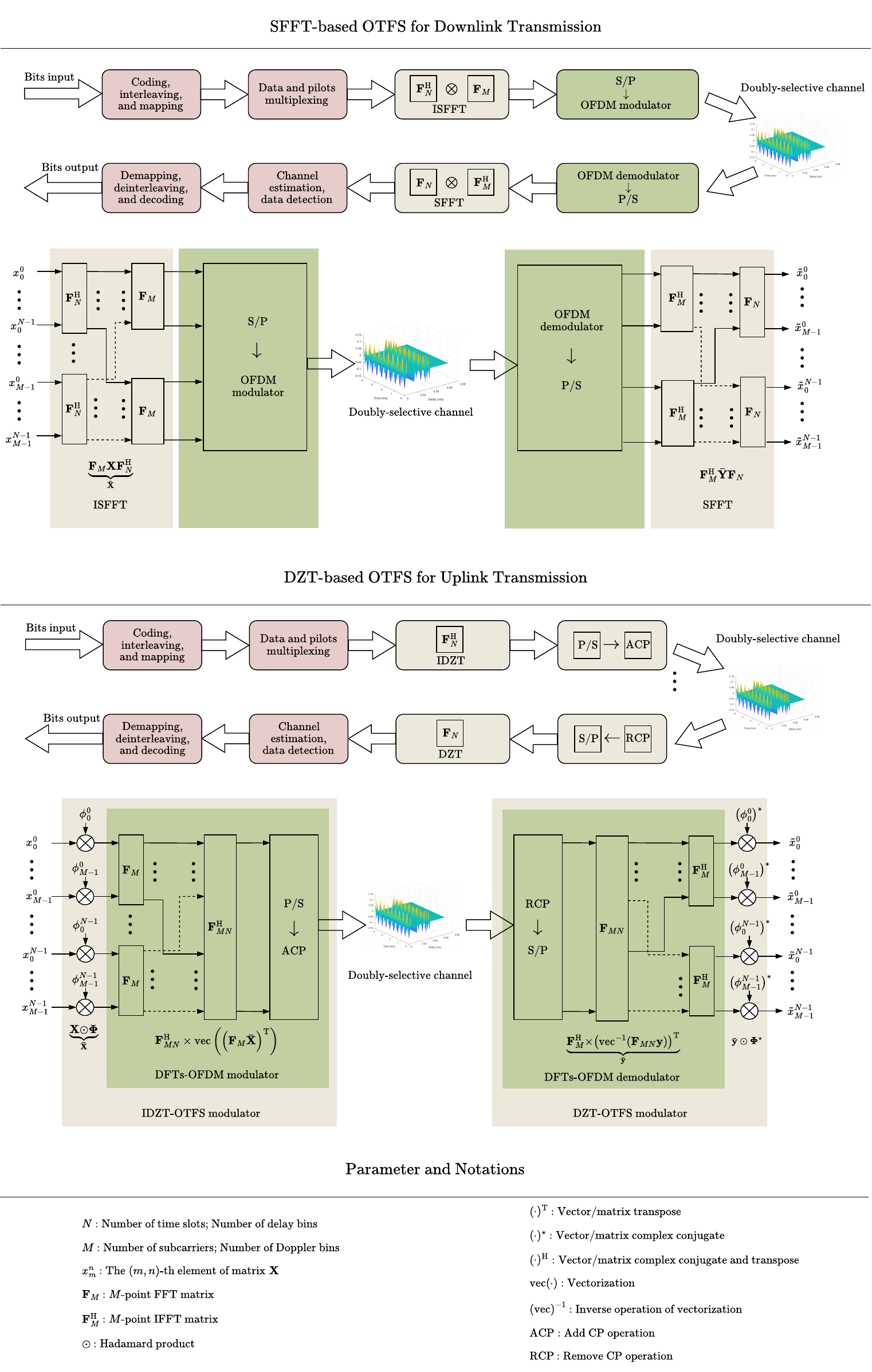}
    \caption{OTFS transceivers for both downlink (upper-figure) and uplink (lower-figure) built upon an OFDM-based architecture.}
    \label{fig:OTFS_implementations}
\end{figure*}

\section{Backward Compatibility of OTFS}
The backward compatibility of OTFS is an important practical issue that currently hinders progress toward its standardization. In this section, we discuss how OTFS can be seamlessly integrated into conventional OFDM-based systems, including both downlink and uplink transmissions.

\subsection{4G and 5G-NR Downlink-Compatible OTFS}

The core idea of OTFS is to directly multiplex information symbols onto a discretized two-dimensional (2D) grid in the DD domain. Specifically, the delay dimension is quantized into bins with a resolution of $\Delta\tau=\frac{1}{M\Delta f}$, where $\Delta f$ is the OFDM subcarrier spacing and $M$ is the total number of subcarriers. Each delay bin is indexed by $l$ with $1\le l\le M$. Similarly, the Doppler dimension is quantized with a resolution of $\Delta\nu=\frac{1}{NT}$, where $T$ is the duration of an OFDM symbol and $N$ is the number of OFDM symbols.\footnote{Here, it is evident that DD waveform demonstrates significantly higher resolution than the TF waveform, whose time and frequency resolutions are $T$ and $\Delta f$, respectively.} This carefully structured discretization allows OTFS to support channels with maximum delay spreads up to $T$ and maximum Doppler shifts up to $\Delta f$,\footnote{Note that a maximum speed of $1000$ km/h can be supported by choosing, for example, a carrier frequency of $4$ GHz and an OFDM subcarrier spacing $\Delta f$ of approximately $4$ kHz.} making it suitable for practically relevant underspread channels. Since physical propagation occurs in the time domain, the OTFS transceiver is designed to convert the DD domain signal into the time domain for actual transmission, and to recover the DD domain signals from the received time-domain signal after experiencing the doubly-selective wireless channel.

One of the major implementation approaches, which is particularly compatible with existing OFDM-based downlink standards, is commonly referred to as the symplectic finite Fourier transform (SFFT)-based implementation~\cite{wei2021orthogonal}, as illustrated in Fig.~\ref{fig:OTFS_implementations}. Specifically, the data symbols multiplexed onto the DD domain grid are first mapped into the TF domain leveraging an inverse SFFT (ISFFT). This operation effectively spreads each DD domain symbol across the entire TF domain, ensuring that all DD domain symbols will principally experience the same TF domain doubly-selective channel. As a result, OTFS potentially exploits the full inherent channel diversity, achieving superior error performance compared to conventional TF domain waveforms. The resultant TF domain signal is then passed into a conventional OFDM modulator\footnote{Note that the OFDM modulator employs $M$ subcarriers and internally includes a serial-to-parallel (S/P) conversion block to map the $MN$-length input sequence onto these subcarriers, as depicted in Fig.~\ref{fig:OTFS_implementations}. A corresponding parallel-to-serial (P/S) conversion block is likewise incorporated in the OFDM demodulator.} along with a TF domain pulse shaping filter to generate the time-domain transmit waveform. After passing through the doubly-selective channel, a conventional OFDM demodulator is deployed, yielding discrete TF domain received symbols. Finally, the DD domain received symbols are recovered by applying SFFT to the demodulated TF domain symbols. 

Thus, it has been clearly demonstrated that SFFT-based OTFS can be seamlessly implemented as an overlay on existing OFDM systems for downlink transmissions. Notice that ISFFT and SFFT operations at the transceiver are treated as additional precoding and post processing blocks placed before and after the standard OFDM modulation and demodulation stages. As shown in~\cite{wei2021orthogonal}, ISFFT can be efficiently realized by performing a fast Fourier transform (FFT) along the delay dimension followed by an inverse FFT (IFFT) along the Doppler dimension, and by performing inverse operations at the receiver to realize the SFFT operation. This straightforward implementation enables OTFS to be integrated into the existing OFDM-based downlink systems with minimal hardware modifications. Furthermore, the resulting time-domain OTFS signal requires only a single CP for an entire frame, rather than inserting a CP for each symbol as in conventional OFDM systems. As a result, SFFT-based OTFS not only achieves higher SE than OFDM but also reduces implementation complexity by eliminating the need for multiple CPs.

\subsection{4G and 5G-NR Uplink-Compatible OTFS}
Different from the downlink OTFS framework that adopts the SFFT-based two-stage implementation, an alternative approach that bypasses the overlay of the TF domain processing stage is the Zak transform (ZT)-based implementation~\cite{farhang2024sc}. 
The ZT, originally introduced by J. Zak in solid-state physics and subsequently revisited from a signal processing perspective, has become an important tool for TF analysis. However, it was not utilized as an information modulation technique until the emergence of generalized frequency division multiplexing (GFDM) and OTFS waveforms. In particular, under the discrete Zak transform (DZT)-based OTFS framework illustrated in Fig.~\ref{fig:OTFS_implementations}, the DD domain symbols are directly mapped to the time domain through the inverse DZT (IDZT). The resultant time-domain samples are then transformed into a continuous waveform via applying appropriate filtering and windowing operations. Similarly, the received time-domain signal after matched filters and windows can be directly converted to the DD domain by applying the DZT. Compared with the SFFT-based OTFS implementation, the DZT-based OTFS structure offers lower computational complexity. This makes it particularly appealing for uplink deployment at user equipment, where hardware resources are typically limited.  More importantly, it has been shown in~\cite{li2024fundamentals} that the DZT can be efficiently implemented by adopting FFT operations. Leveraging this property, the DZT-based OTFS transceiver can be realized with further reduced complexity. Specifically, at the transmitter side, the DD domain symbols are first processed by a single $N$-point IFFT along the Doppler dimension, followed by a symbol-wise interleaving. At the receiver, after applying the corresponding deinterleaving operation, an $N$-point FFT along the Doppler dimension is performed to recover the DD domain symbols~\cite{farhang2017low}.

Alternatively, the DZT-based OTFS transmission can be equivalently realized as a discrete Fourier transform-spread OFDM (DFT-s-OFDM) structure with additional phase rotations~\cite{farhang2024sc}. As shown in Fig.~\ref{fig:OTFS_implementations}, considering the DZT-based OTFS transmission with $N$ time slots and $M$ subcarriers, the DD domain symbols $\mathbf{X}_{\mathrm{DD}}\in\mathbb{C}^{M\times N}$ are first multiplied element-wise by a phase compensation matrix $\boldsymbol{\Phi}\in\mathbb{C}^{M\times N}$, whose $(m,n)$-th entry is given as $e^{-j\frac{2\pi}{MN}mn}$, where $0\le m\le M-1$ and $0\le n\le N-1$. The resultant matrix is then left multiplied by an $M$-point DFT along the delay dimension. The output is subsequently transposed and vectorized, i.e., symbol-wise interleaving, followed by an $MN$-point inverse DFT (IDFT) to generate the final time-domain OTFS symbols~\cite{farhang2024sc}. Similarly, the receiver processing can be implemented by performing the inverse sequence of operations applied at the transmitter~\cite{farhang2024sc}.

It is worth highlighting that the presented approach reveals a close relationship between DZT-based OTFS and the DFT-s-OFDM waveform, which is employed for uplink transmission in 4G and 5G-NR systems due to its ability to achieve a low peak-to-average power ratio (PAPR). The primary difference between these two waveforms lies in the introduction of additional linear phase rotations applied prior to the DFT precoder at the transmitter and after the IDFT post-processing unit at the receiver~\cite{farhang2024sc}. This property is particularly promising, as it ensures strong backward compatibility with existing 4G and 5G-NR networks, while enabling low-cost OTFS implementation from the manufacturer's perspective. Moreover, the DZT-based OTFS waveform inherently exhibits favorable low-PAPR characteristics, positioning it as a naturally suitable candidate for uplink transmission.

\subsection{Connections Between Downlink and Uplink OTFS}

{In fact, both SFFT-based and DZT-based OTFS are essentially approximations of the continuous DD waveform based on the ZT. Although differences exist between these two schemes, they can yield similar input–output relations depending on the specific choices of underlying filters and windows~\cite{li2024fundamentals}. In particular, the SFFT-based OTFS employs a TF-domain pulse-shaping filter for carrying the information symbols, where bi-orthogonal pulses are desirable to effectively suppress TF-domain interference, thereby producing a well-structured channel matrix in the DD domain. By contrast, the DZT-based OTFS utilizes a time-domain pulse-shaping filter, where the pulses are designed to be orthogonal with respect to delay resolution to achieve approximate DD orthogonality while also controlling the bandwidth. Furthermore, window design is applicable only in the TF domain of the SFFT-based OTFS framework. For example, applying a Dolph–Chebyshev (DC) window can enhance the effective channel sparsity by suppressing the sidelobes of the channel response, thereby reducing the fractional Doppler components in the DD domain~\cite{wei2021transmitter}. }

\subsection{Delay-Doppler Multiple Access}
{

In conventional orthogonal frequency-division multiple access (OFDMA), users are allocated distinct subsets of subcarriers. Due to the multiplicative nature of the TF-domain channel, orthogonality among users is preserved, thereby avoiding inter-user interference. However, DD multiple access inherently encounters interference from different users due to the twisted convolution property of DD communications. As a result, interference management becomes a central design objective for DD multiple access. One straightforward solution is to allocate distinct resource blocks in the DD domain to different users, with sufficiently large guard spaces between adjacent users to avoid inter-user interference~\cite{hong2022delay}. However, this approach comes at the expense of significant resource overhead. Alternatively, employing advanced interference-cancellation mechanisms tailored to DD communications offers a promising direction, the details of which will be discussed in the following section.
}

\section{Complexity-Scalable Multiuser MIMO-OTFS Systems}

\begin{figure*}[t!]
    \centering
    \includegraphics[scale=0.6]{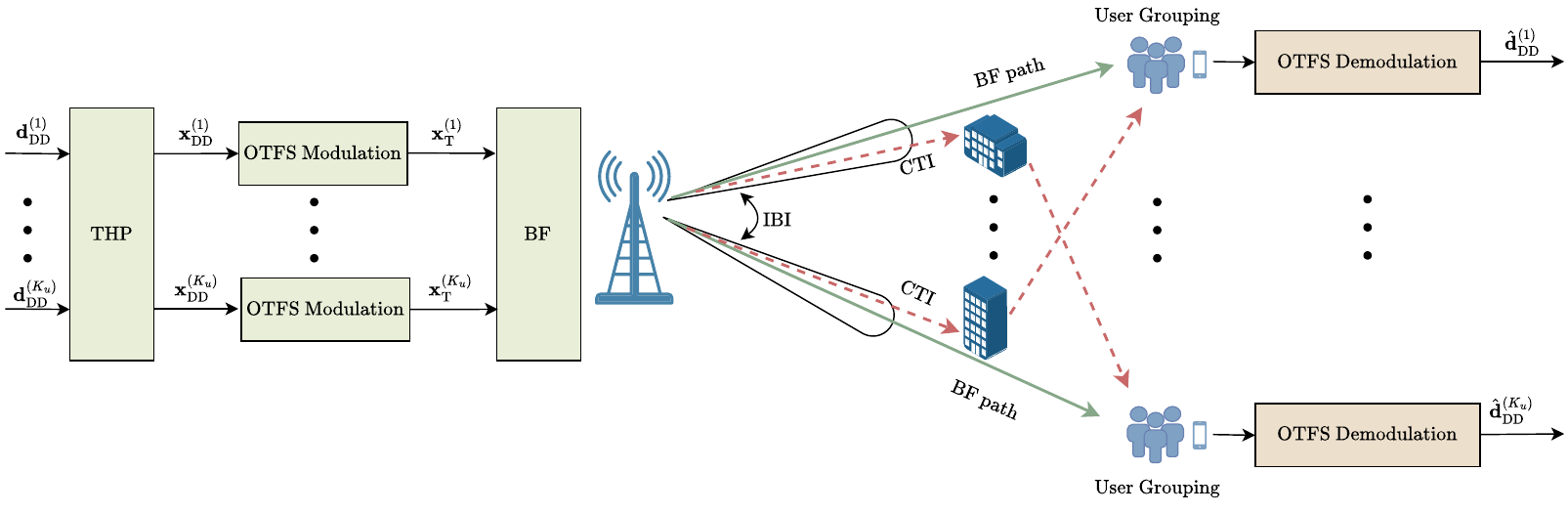}
    \caption{The block diagram and interference pattern of the considered THP-based downlink MU-MIMO-OTFS transmissions.}
    \label{fig:DD_THP_TxRx}
\end{figure*}
MIMO technology has become a key enabler for enhancing system capacity and reliability in current 5G-NR wireless networks, and it will be integral to future network generations. Hence, exploring the integration of OTFS in multiuser MIMO systems, i.e., MU-MIMO-OTFS, is essential for achieving Doppler spread resilience and high data-rate communications. However, the design of efficient MU-MIMO-OTFS remains challenging. Unlike OFDM-based MIMO, which modulates symbols independently on each subcarrier and handles interference only from narrowband equivalent channels, OTFS-based MIMO operates over wideband, doubly selective channels, where interference must be managed globally across the entire DD domain. This leads to a fundamentally different interference pattern compared with conventional OFDM-based MIMO systems.
Consequently, conventional MIMO designs are generally not applicable to OTFS. For instance, standard linear precoders, such as zero forcing (ZF), require inversion of the effective MIMO channel matrix. However, in the DD domain, this effective channel matrix becomes exceedingly large, with its dimensions scaling with the product of the number of delay bins, Doppler bins, and spatial resource elements. This results in prohibitively high computational complexity for practical MU-MIMO-OTFS systems. While OFDM-based MIMO benefits from implementation simplicity under the assumption of subcarrier orthogonality, this advantage deteriorates rapidly in high-Doppler scenarios. In contrast, as we will demonstrate, MU-MIMO-OTFS can be implemented with near-linear computational complexity by leveraging the separability of the DD-domain channel and the inherent benefits of global interference management.

To better illustrate the structure of MU-MIMO-OTFS and the corresponding interference pattern, let us consider a downlink MU-MIMO-OTFS transmission, where a base station (BS) equipped with $N_{\mathrm{t}}$ antennas serves $K_{u}$ single-antenna users. As shown in Fig.~\ref{fig:DD_THP_TxRx}, the DD domain information symbols for the $u$-th user, denoted by $\mathbf{d}^{(u)}_{\mathrm{DD}}$, for $1\le u\le K_u$, are first processed via the DD domain Tomlinson–Harashima precoding (THP) scheme, whose details will be discussed later. The precoded symbols $\mathbf{x}^{(u)}_{\mathrm{DD}}$ are then transformed to the time domain through OTFS modulation, yielding $\mathbf{x}^{(u)}_{\mathrm{T}}$. The resultant time-domain signal is subsequently passed through a conventional beamforming (BF) stage,  generating the downlink transmit signals $\mathbf{z}_{n}$ for $1\le n\le N_{\rm{t}}$. We assume that the BS deploys a uniform linear array (ULA). To capture the combined effect of beamforming and the spatial steering vector for the $u$-th user, we define an \emph{effective spatial-domain channel vector} $\mathbf{g}^{(u)}_p$, where $0\le p\le P-1$. After propagating through the doubly-selective time-domain channel, each user applies OTFS demodulation to recover the received signal $\hat{\mathbf{d}}^{(u)}_{\mathrm{DD}}$ in the DD domain.

As illustrated in Fig.~\ref{fig:DD_THP_TxRx}, the BS generates multiple beams, each directed toward the dominant propagation path of each user, referred to as the BF path, while all remaining paths of each user are termed non-BF paths. As a result, except for the desired signal that carries the intended information for the target user and is transmitted via its BF path, three different types of interference emerge:
\begin{itemize}
    \item Multipath self-interference  (MPSI): The self-interference caused by the desired user's non-BF paths due to the multipath propagation effects.
    \item Inter-beam interference (IBI): The interference results from beams directed towards other users inadvertently leaking into the beam direction of the desired user. 
    \item Crosstalk interference (CTI):  The interference 
    occurs when a beam intended for the BF path of one user overlaps with and thus interferes with the unintended non-BF path of another user. In other words, a beamformed signal is inadvertently transmitted towards the non-BF path of an unintended user.  
\end{itemize}
While we identify three types of interference with distinct physical interpretations, it is important to note that not all of them contribute significantly to the received signal. In practice, user grouping strategies are typically employed at the BS to mitigate severe interference among users. Specifically, users with diverse spatial characteristics, such as angles of departure (AoDs), are grouped together. Consequently, by leveraging appropriate user grouping and well-designed beam codebooks, the effects of IBI and MPSI can be substantially reduced.

\begin{figure}[t]
    \centering
    \includegraphics[scale=0.5]{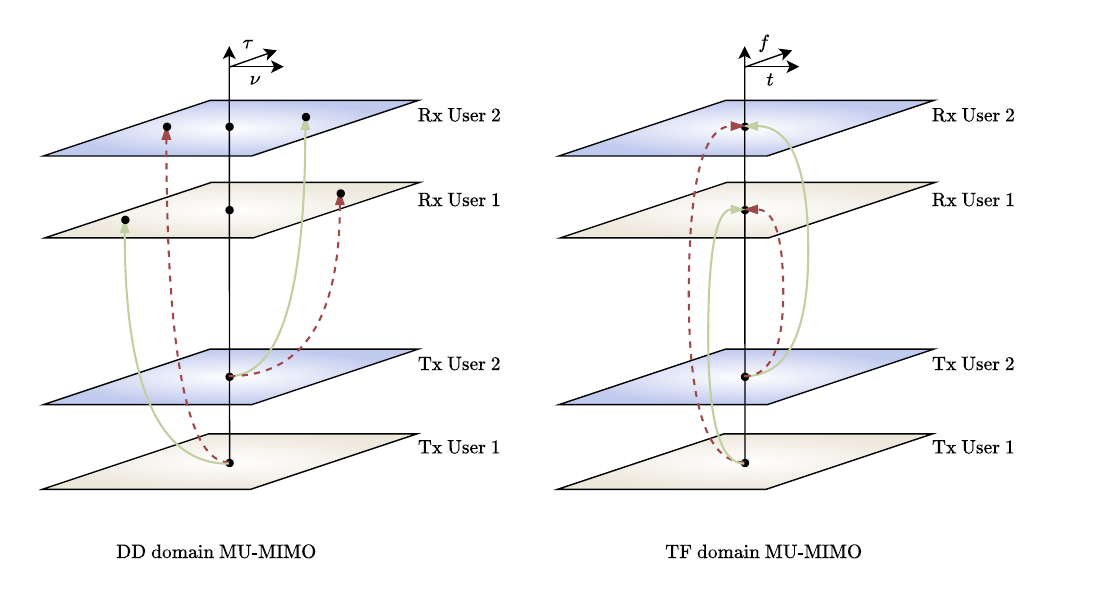}
    \caption{Interference pattern comparison between MU-MIMO-OTFS and MU-MIMO-OFDM.}
    \label{fig:interference_pattern}
\end{figure}

However, CTI remains a practical challenge, as the BF path of one user can inadvertently overlap with a non-BF path of another user. In such cases, the interference received through the non-BF path can be significant due to the strong beamforming gain, even though the non-BF path itself has a relatively low channel gain. Interestingly, since resolvable paths from different users typically do not share identical delay, Doppler, and angular characteristics, the interference observed in each DD-domain received symbol originates from different DD-grid locations of other users. In other words, MU-MIMO-OTFS naturally separates the desired signal and the interference originating from the same transmitted DD grid into distinct DD grids at the receiver, as illustrated on the left-hand side of Fig.~\ref{fig:interference_pattern}. This behavior stands in sharp contrast to OFDM, where interference affecting a received symbol in the TF domain often arises from the same TF grid across different users, as shown on the right-hand side of Fig.~\ref{fig:interference_pattern}. The underlying reason for this difference lies in the convolutional nature of the DD-domain channel, in contrast to the multiplicative nature of the TF-domain channel. Indeed, the separability of the DD-domain channel allows us to precisely identify the origins of interference for each received symbol, thereby enabling the design of reduced-complexity precoders for MU-MIMO-OTFS transmissions.

THP is a classical nonlinear precoding technique that is particularly effective for mitigating the CTI described above~\cite{li2023delay}. The core principle of THP is to pre-cancel interference at the transmitter based on a known interference pattern, which can be fully determined from the available channel state information (CSI). However, directly applying THP to MU-MIMO-OTFS systems requires a QR decomposition, which is computationally prohibitive because the effective channel matrix in MU-MIMO-OTFS systems is extremely large, with dimensions scaling across delay, Doppler, and spatial domains. In contrast, implementing THP in the DD domain enables an almost linear-complexity precoder. This improvement arises from the fact that DD-domain channel paths are distinguishable in at least one of the delay, Doppler, or angular dimensions. Such separability allows for precise identification of interference sources in the DD domain. Consequently, by strategically placing known symbols at selected positions, it becomes possible to break the interference cycle in the DD domain, allowing all interference to be cancelled in a symbol-by-symbol manner~\cite{li2023delay}.

\section{Delay-Doppler Integrated Sensing and Communications}

As aforementioned, IMT-2030 envisions sensing as an integral new capability of 6G networks to enable emerging use cases and application trends~\cite{ITU2023}. To realize this vision and fulfill the stringent sensing-related KPIs of 6G, ISAC plays a vital role by enhancing performance and reducing the overall cost of both systems. Among various ISAC schemes, DD-ISAC has recently attracted significant attention due to its inherent advantages, such as its unified transceiver architecture. Conventional ISAC schemes, such as OFDM-ISAC, typically require additional and distinct modules to realize communication and sensing functionalities simultaneously. However, this approach deviates from the fundamental ISAC objective of seamless hardware and software integration. In contrast, beyond its communication functionality, DD-ISAC offers a more integrated sensing solution by exploiting the natural overlap between the range–Doppler matrix computation and the demodulation process of OTFS signals. Specifically, the range–Doppler matrix is typically constructed by applying a DFT along the slow-time dimension of the fast-time/slow-time matrix, which characterizes the target response along delay and Doppler after matched filtering and sampling. Based on the earlier discussions, it can be observed that the demodulation procedure of DZT-based OTFS is mathematically identical to the range–Doppler computation in radar sensing, where the time-domain signal after matched filtering is transformed into the DD domain via a DFT. The only distinction lies in their primary objectives: range–Doppler processing is used to extract sensing parameters, whereas OTFS demodulation aims to recover information symbols.

Due to these features, DD-ISAC holds great potential for enabling tightly integrated sensing and communication sub-systems. For example, in sensing-assisted communications, sensing results can be directly used as \emph{a priori} channel information, thereby reducing communication overhead. However, in conventional OFDM-ISAC systems, additional transformations are required to convert sensing outputs into a form suitable for communication tasks. In contrast, DD-ISAC systems enable direct exploitation of sensing results, not only because OTFS modulation inherently operates in the DD domain, but also due to DD reciprocity,\footnote{ Unlike the channel reciprocity in time-division duplex (TDD) systems, DD reciprocity is defined within the stationarity region, denoting that the underlying delay and Doppler shifts remain invariant between uplink and downlink transmissions. Although these two notions of ``reciprocity'' may coexist in practice, one should focus on exploiting channel reciprocity in either the time domain or the DD domain for system design, since they often require significantly different system setups.} which allows sensing-derived channel knowledge to be directly applied for precoding at the transmitter, as long as the transmissions occur within the same stationarity region. Similarly, in communication-assisted sensing, the channel estimates obtained in the DD domain can be directly used as coarse sensing information, eliminating the need for TF-to-DD conversion required in conventional OFDM-ISAC systems. More importantly, OFDM employs TF-domain pilots placed at selected positions to acquire CSI, while the CSI at non-pilot positions must be obtained through interpolation—whose accuracy degrades significantly in doubly selective channels. In contrast, OTFS employs DD-domain pilots that spread across the entire TF domain, enabling the acquisition of more accurate and robust CSI for sensing assistance.

Furthermore, DD-ISAC achieves improved sensing performance by exploiting the unique properties of DD domain basis functions. In particular, the DD domain basis functions, which carry the DD domain symbols, including pilots, exhibit a direct relationship with the sensing ambiguity function. 
As demonstrated in~\cite{li2024fundamentals}, this connection can be simply characterized by a twisted-correlation, indicating that the design of DD domain basis functions significantly impacts the ambiguity function, which is a fundamental indicator of the sensing performance. Note that the ambiguity function is conceptually connected to the widely adopted Cram$\Acute{\text{e}}$r-Rao Bound (CRB) in many sensing applications (see~\cite{gaudio2020effectiveness} and the references therein). For example, in the single target scenario, the CRB value is dominated by the relative height of the mainlobe to the sidelobes of the ambiguity function in the low signal-to-noise ratio (SNR) regime, which reflects the likelihood of having “large errors” dominated by outliers. Conversely, in the high SNR regime, the CRB value is dominated by the curvature of the ambiguity function’s mainlobe, which is inherently related to the Fisher information matrix. In this paper, however, we focus on the underexplored ambiguity function to highlight the interesting insights arising from its intrinsic relationship with the underlying basis function. {Furthermore, the ambiguity function is a fundamental tool in pulse-shaping filter design, offering the potential for a unified waveform design framework that integrates pulse shaping and basis function design.}

Specifically, unlike the TF-domain basis functions adopted in OFDM, which suffer from reduced pulse localization due to Heisenberg’s uncertainty principle, DD-domain basis functions exhibit excellent localization, featuring sharp peaks within the fundamental rectangle, defined as $[\tau,\nu]\in[0,T]\times[0,1/T]$. Notably, the DD-domain basis functions maintain a globally quasi-periodic structure, thereby remaining consistent with Heisenberg’s uncertainty principle. Consequently, it is possible to achieve a sharply peaked ambiguity function by carefully designing the DD basis functions through an appropriate selection of $T$ based on the underlying channel conditions, ensuring that the basis functions are well concentrated. Although constructing ideal DD-domain basis functions would require infinite time and frequency support, and is therefore impractical for real-world implementation, it has been demonstrated in~\cite{li2024fundamentals} that truncating localized DD basis functions induces a transition from DD localization to DD orthogonality with respect to the delay and Doppler resolutions, enabling practical implementation alternatives.\footnote{Note that DD localization implies that an impulse in the DD domain behaves as an ideal delta function without sidelobes overlapping neighboring symbols. In contrast, DD orthogonality allows DD-domain pulses to overlap without introducing interference, as they are separated at integer multiples of the orthogonality period. However, DD orthogonality cannot fully eliminate fractional delay and Doppler effects, since these impairments arise from finite DD-domain resolution rather than from the lack of pulse orthogonality.} This truncation is achieved through time and frequency windowing, where the choice of window shape directly influences the ratio between the mainlobe and sidelobes of the ambiguity function. It has been shown in~\cite{li2024fundamentals} that employing truncated periodic window functions enables the truncated basis functions to achieve sufficient orthogonality, thereby yielding favorable ambiguity function characteristics for accurate sensing functionality.

\section{Performance Evaluation of OTFS}

\begin{figure}[h]
	\centering  
    \subfigure[]{{\includegraphics[scale=0.58]{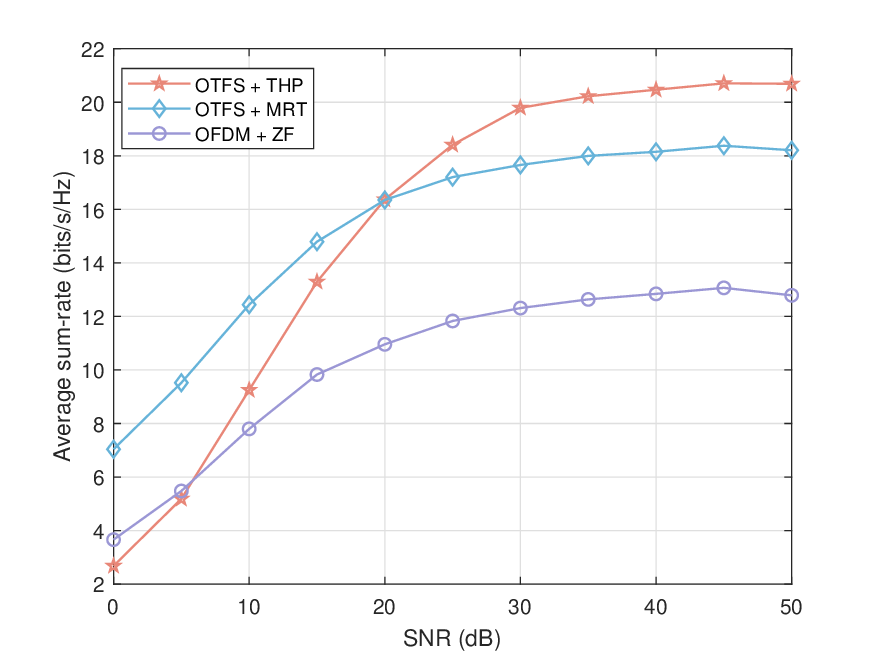}}}
	\subfigure[]{{\includegraphics[scale=0.4]{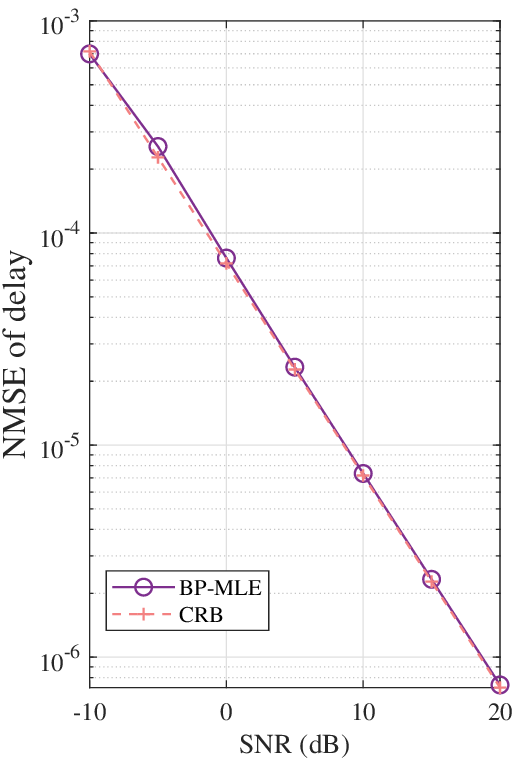}}}
    \subfigure[]{{\includegraphics[scale=0.4]{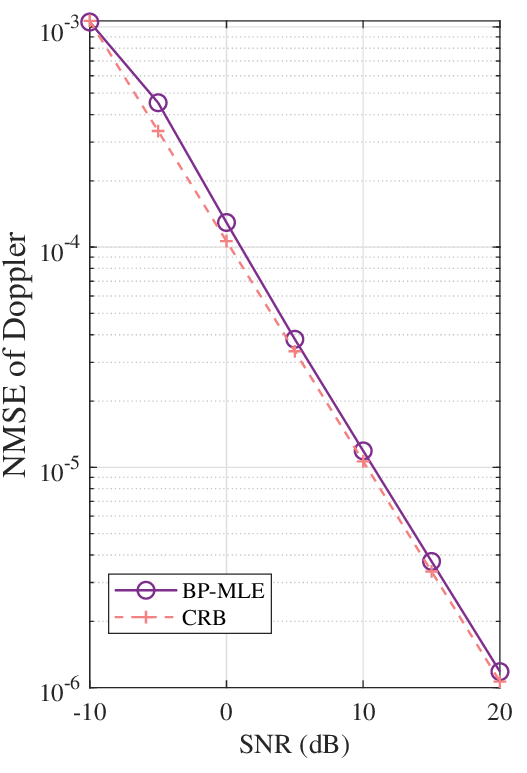}}}
    \vspace{-2mm}
    \caption{Average sum-spectral efficiency and sensing performance of the OTFS waveform:
(a) Average sum-SE comparison in MIMO systems using different waveforms and precoding schemes, including OTFS-THP, OTFS-MRT, and OFDM-ZF;
(b) Doppler sensing performance comparison between OTFS and the CRB;
(c) Delay sensing performance comparison between OTFS and the CRB.} 
        \label{fig:comparisons}
\end{figure} 

In this section, we demonstrate the superiority of OTFS over conventional OFDM. Specifically, we consider a MU-MIMO-OTFS transmission scenario in which a BS equipped with $N_{\mathrm{t}} = 8$ antennas serves four users. The OTFS frame size is configured with parameters $N = 16$ and $M = 32$. The channel’s maximum delay and Doppler indices are set to $l_{\max} = 5$ and $k_{\max} = 7$, respectively, with all indices assumed to be integer values. Each user experiences three fading paths, whose coefficients are generated based on an exponentially decaying power delay profile with a path-loss exponent of $2.76$. The THP-based transmission scheme described in~\cite{li2023delay} is adopted, and all three types of interference (MPSI, IBI, and CTI) are taken into account.

The average sum-SE performance under these settings is illustrated in Fig.~\ref{fig:comparisons}(a). It can be observed that the OTFS waveform with THP achieves a SE comparable to that of ZF-precoded OFDM at low SNRs, although it initially performs below maximum ratio transmission (MRT)-precoded OTFS. The rate loss of THP at low SNR is primarily attributed to the so-called “modulo loss,” which arises from the modulo operation performed at the receiver. However, this loss diminishes rapidly as the SNR increases, as shown in the figure. Notably, at high SNRs, the proposed OTFS–THP scheme outperforms all existing schemes in terms of sum-SE. Additionally, the sum-SE saturates when the SNR exceeds $30$ dB, primarily due to the effects of MPSI and IBI, which depend only on the transmit and channel power. Since these two interference components are independent of noise power, increasing the SNR cannot mitigate their impact, resulting in a noticeable error floor. Moreover, although OTFS–THP requires inserting known symbols in the DD domain—leading to some signaling overhead—this overhead can be reduced through appropriate overhead-reduction strategies~\cite{li2023delay}.

For sensing performance evaluation, we consider a challenging scenario with two closely spaced targets in both delay and Doppler dimensions, where conventional OFDM-based sensing typically fails to provide reliable resolution. Specifically, a strong target is located at $l_{1} = 4.1$, $k_{1} = 5.2$, while a weaker target is positioned at $l_{2} = 4.3$, $k_{2} = 5.7$, with the weaker target’s reflectivity being $3$ dB lower than that of the strong target. To address this challenge, we adopt a backpropagation-based maximum likelihood estimation (BP-MLE) approach for both delay and Doppler sensing~\cite{hu2025novel}. The resulting normalized mean squared error (NMSE) performance is presented in Fig.~\ref{fig:comparisons}(b) and Fig.~\ref{fig:comparisons}(c), respectively, alongside the CRB, which represents the theoretical bound for unbiased estimation. The results clearly demonstrate that OTFS-based sensing achieves delay and Doppler estimation accuracies that closely approach the CRB, even in this challenging scenario, thereby highlighting its robustness and effectiveness for high-resolution sensing.

\section{Opportunities and Challenges}

Although OTFS modulation has been studied in the existing literature, it is important to note that research in this area is still far from comprehensive. Several critical issues remain insufficiently explored, and further investigation is required to enable practical deployment and performance optimization. In the following, we highlight several promising research directions and open problems associated with OTFS modulation that warrant future exploration.

\subsection{Pathways towards Standardization for DD waveforms}

As previously discussed, OTFS can be effectively implemented on top of existing downlink OFDM and uplink DFT-s-OFDM systems, which are already standardized in current 4G and 5G-NR networks. This backward compatibility offers strong potential for OTFS to be integrated into future 3rd Generation Partnership Project (3GPP) releases as part of the evolutionary path toward 6G. However, standardization efforts for OTFS and general DD waveforms are still in their early stages, particularly in terms of prototyping and large-scale validation. Furthermore, low-complexity receiver designs are crucial for the practical standardization of DD waveforms, and may require adjustments to existing receiver architectures to fully exploit the advantages of DD-domain processing. As such, advancing the standardization of DD waveforms, while maintaining strong backward compatibility with existing OFDM systems, represents a promising and impactful research direction.

\subsection{Theoretical Foundations of DD Sensing}

Since communication remains the core service of 6G, DD-ISAC systems must maximize the utilization of limited available resources to ensure sufficient communication throughput. However, this requirement inevitably leads to the use of random communication signals for sensing tasks, whereas traditional sensing typically relies on deterministic waveforms. Consequently, the inherent randomness of information-bearing symbols can significantly affect sensing accuracy, and the theoretical analysis and fundamental limits of this challenge remain open research questions. Understanding how DD-ISAC signals carrying random data payloads influence sensing performance is therefore crucial for characterizing the non-trivial fundamental trade-off between communication and sensing in the DD domain.

\subsection{Ultra Reliable and Low Latency-Coded DD Communications}
As previously discussed, OTFS distributes information symbols across the entire TF domain, typically requiring the reception of the complete OTFS frame before detection. This inherently results in increased latency compared to conventional TF-domain waveforms, posing challenges for meeting the stringent latency KPIs of 6G. Although reducing the OTFS frame duration can alleviate latency, the resulting performance often remains suboptimal. Therefore, it is necessary to develop transceiver architectures that can detect DD-domain modulated information symbols with sufficient accuracy using only a portion of the received signal, thereby reducing latency. {A potential approach is to adopt a cross-domain framework, which leverages the channel properties of the DD domain while retaining the conventional TF domain transmission to satisfy latency requirements.} In this context, channel coding becomes critical for improving error performance. However, coding schemes specifically tailored to DD communications remain underexplored. Thus, designing coding strategies that not only exploit DD-domain channel characteristics but also enable reliable decoding under partial reception represents an important direction for future research.

\subsection{Delay-Doppler Radiomap}
Radiomap channel prediction has recently attracted considerable attention due to its capability to reduce pilot overhead in communication systems. By leveraging advanced machine learning networks, radiomap technology constructs a geographical representation of the signal power spectral density as a function of location, time, and frequency, providing valuable \emph{a priori} channel knowledge to assist communication tasks. However, most existing radiomap systems do not explicitly account for mobility-induced variations in channel behavior. As a result, the \emph{a priori} channel knowledge provided by conventional radiomaps becomes less effective in high-mobility scenarios, where TF-domain channels vary rapidly. In contrast, the DD-domain channel representation directly captures the underlying physical scattering geometry of the environment, which remains relatively stable over longer time intervals. Consequently, constructing a radiomap in the DD domain is expected to require substantially less training overhead and to offer more robust channel prediction compared with conventional TF-domain radiomap methods. Nevertheless, this direction remains in its early stages, and further investigation is required to fully unleash the potential of DD-domain radiomap techniques.

\section{Conclusion}
OTFS has emerged as a highly promising waveform candidate for satisfying the stringent requirements outlined in the 6G KPIs, thanks to its inherent Doppler resilience and superior sensing capability. In this article, we examined the fundamental advantages of DD-domain waveforms and explored the practical implementation of OTFS within existing OFDM-based systems. We began by providing an in-depth discussion of the DD-domain channel representation, thereby unveiling the underlying rationale behind the OTFS technique. Subsequently, we demonstrated the compatibility of OTFS with current OFDM-based downlink and uplink systems, highlighting clear pathways for practical deployment. Furthermore, we presented a detailed study of a practical MU-MIMO-OTFS precoding scheme, whose implementation complexity scales linearly with the channel dimensions, making it suitable for large-scale systems. The superiority of DD-ISAC was also emphasized, particularly through its unified transceiver design and its ability to achieve favorable ambiguity-function characteristics via carefully designed basis functions. Finally, we identified several key challenges and outlined promising research directions for advancing OTFS technology toward efficient, scalable, and robust 6G communications.

\bibliographystyle{IEEEtran}
\bibliography{ref}

\begin{IEEEbiographynophoto}
{Mingcheng Nie} received the B.E. degree and the M.Phil. degree in electrical engineering from the University of New South Wales (UNSW), Australia, in 2021 and 2024, respectively. He is currently pursuing the Ph.D. degree with the school of electrical and computer engineering, the University of Sydney (USYD), Australia. His research interests include waveform design, deep learning, and resource allocation.
\end{IEEEbiographynophoto}

\begin{IEEEbiographynophoto}
{Ruoxi Chong} (Member, IEEE) obtained her Ph.D. degree from Queen's University Belfast, U.K. in 2025. She is currently a research fellow with the Center for Wireless Innovation (CWI), Queen's University Belfast, U.K. She received the B.Sci. degree in Optoelectronic Information Science and Engineering from Beijing Jiaotong University, Beijing, China, and the M.Eng. degree in Electrical Engineering (Telecommunications) from the University of New South Wales (UNSW), Sydney, Australia, in 2019 and 2021, respectively. Her main research interests include MIMO systems, ISAC, and waveform design.
\end{IEEEbiographynophoto}


\begin{IEEEbiographynophoto}
{Shuangyang Li} (Member, IEEE) is a Marie Skłodowska-Curie Actions (MSCA) research fellow at Technische Universität Berlin. He is the recipient of the best young researcher award 2024 from the IEEE ComSoc EMEA region, and the European Research Council (ERC) starting grant 2025. He received Best Paper Awards from IEEE ICC 2023, IEEE GlobeCom 2025, and IEEE/CIC ICCC 2025, and the Best Workshop Paper Award from IEEE WCNC 2023. His research interests include signal processing, applied information theory, and their applications to communication systems.
\end{IEEEbiographynophoto}

\begin{IEEEbiographynophoto}
    {Arman Farhang} (Senior Member, IEEE) received the Ph.D. degree from Trinity College Dublin, Ireland, in 2016. He is currently an Assistant Professor with the Electronic and Electrical Engineering Department at Trinity College Dublin and a Principal Investigator on multiple research projects. He is a member of Research Ireland centres CONNECT, ADVANCE-CRT, and ARC Hub in ICT, where he leads research on waveform design and multi-antenna systems. His research interests include wireless communications, advanced modulation formats, synchronization and channel estimation, and multiuser and multi-antenna communications.
\end{IEEEbiographynophoto}

\begin{IEEEbiographynophoto}
    {Fabian Göttsch} is a Research Engineer with Massive Beams. He received his M.Sc. and Ph.D. in Electrical Engineering from the Technical University of Berlin in 2020 and 2024, respectively. His research interests include the modeling, optimization, and experimental evaluation of wireless networks, with a focus on Open RAN and next-generation multi-antenna systems.
\end{IEEEbiographynophoto}

\begin{IEEEbiographynophoto}
    {Derrick Wing Kwan Ng} (Fellow, IEEE) received
his Ph.D. degree from the University of British
Columbia in 2012. He is now working as an Associate Professor at the University of New South Wales, Sydney,
Australia. His research interests include convex and
non-convex optimization, physical layer security,
wireless information and power transfer, and green
(energy-efficient) wireless communications.
\end{IEEEbiographynophoto}

\begin{IEEEbiographynophoto}
{Michail Matthaiou}(Fellow, IEEE) obtained his Ph.D. degree from the University of Edinburgh, U.K. in 2008. 
He is currently a Professor of Communications Engineering and Signal Processing and Deputy Director of the Centre for Wireless Innovation (CWI) at Queen’s University Belfast, U.K. He is also an Eminent Scholar at the Kyung Hee University, Republic of Korea. 
He currently holds the ERC Consolidator Grant BEATRICE (2021-2026) focused on the interface between information and electromagnetic theories.
His research interests span signal processing for wireless communications, beyond massive MIMO, reflecting intelligent surfaces, mm-wave/THz systems and AI-empowered communications.

\end{IEEEbiographynophoto}

\begin{IEEEbiographynophoto}
    {Yonghui Li} (M’04-SM’09-F19) received his PhD degree in November 2002 from Beijing University of Aeronautics and Astronautics. Since 2003, he has been with the Centre of Excellence in Telecommunications, the University of Sydney, Australia. He is now a Professor and Director of Centre for Telecommunications and IoT at the University of Sydney. He is the recipient of the Australia Research Council (ARC) Queen Elizabeth II Fellowship in 2008, ARC Future Fellowship in 2012 and ARC Industry Laureate Fellowship in 2025. He is listed as a Clarivate highly cited researcher. He is a Fellow of IEEE.
 

\end{IEEEbiographynophoto}

\vfill

\end{document}